\documentstyle[twoside,epsf,12pt]{article}

\newcommand{\noq}{|\emptyset\rag}
\newcommand{\psd}{\psi^{\dagger}}
\newcommand{\psdp}{\psi'^{\dagger}}
\newcommand{\bpi}{\pi}

\makeatletter
\@addtoreset{equation}{section}
\makeatother

\newcommand{\bibi}{\bibitem}

\def\a{\alpha}
\def\b{\beta}
\def\c{\chi}

\def\d{\delta}
\def\e{\epsilon}   
\def\g{\gamma}
\def\h{\eta}

\def\j{\psi}
\def\l{\lambda}
\def\m{\mu}
\def\hm{\hat{\mu}}
\def\n{\nu}
\def\o{\omega}
\def\s{\sigma}    
\def\t{\tau}

\def\x{\xi}

\def\G{\Gamma}

\def\O{\Omega}

\def\X{\Xi}

\def\jb{\overline{\j}}
\def\cb{\overline{\c}}

\def\ad{a^{+}}
\def\bd{b^{+}}
\def\Th{\hat{T}}
\def\CH{{\cal H}}

\def\CHT{\widetilde{\CH}}
\def\Qh{\hat{Q}}

\def\Oh{\hat{\O}}
\def\Ohd{\hat{\O}^{\dagger}}

\newcommand{\jp}{\psi^{\prime}}
\newcommand{\jbp}{\overline{\j}^{\prime}}

\newcommand{\half}{\mbox{{\normalsize $\frac{1}{2}$}} }

\newcommand{\Tr}{\mbox{Tr\,}}
\newcommand{\ra}{\rightarrow}

\newcommand{\Lm}{\Lambda}

\newcommand{\al}{\alpha}

\newcommand{\bt}{\beta}
\newcommand{\lag}{\langle}
\newcommand{\rag}{\rangle}
\newcommand{\gm}{\gamma}

\newcommand{\dl}{\delta}
\newcommand{\ep}{\varepsilon}

\newcommand{\et}{\eta}
\newcommand{\th}{\theta}

\newcommand{\sg}{\sigma}
\newcommand{\ta}{\tau}

\newcommand{\ps}{\psi}
\newcommand{\om}{\omega}

\newcommand{\nnn}{\nonumber \\}

\newcommand{\Ps}{\Psi}

\newcommand{\Om}{\Omega}

\newcommand{\psb}{\overline{\ps}}

\newcommand{\dmu}{\partial_{\mu}}

\newcommand{\ah}{\hat{a}}
\newcommand{\bh}{\hat{b}}

\newcommand{\jh}{\hat{\j}}
\newcommand{\ahd}{\hat{a}^{\dagger}}
\newcommand{\bhd}{\hat{b}^{\dagger}}

\newcommand{\jhd}{\hat{\j}^{\dagger}}

\newcommand{\hmu}{\hat{\mu}}

\newcommand{\aleq}{\mbox{}^{\textstyle <}_{\textstyle\sim}}
\newcommand{\ageq}{\mbox{}^{\textstyle >}_{\textstyle\sim}}

\newcommand{\be}{\begin{equation}}
\newcommand{\ee}{\end{equation}}
\newcommand{\bea}{\begin{eqnarray}}
\newcommand{\eea}{\end{eqnarray}}
\newcommand{\eq}{\ref}
\newcommand{\beq}{\begin{equation}}
\newcommand{\eeq}{\end{equation}}
\newcommand{\cc}{\cite}
\newcommand{\lb}{\label}

\def \3{\ss}

\def\footnoteitem(#1)#2{
\begin{list}{#1}{\labelwidth4.0mm \leftmargin7.0mm
\labelsep2.5mm \rightmargin7.0mm \parsep0.5ex plus0.2ex minus0.1ex
\itemsep0ex plus0.2ex }
\item #2
\end{list}
}

\begin{document}


\headsep=0.0cm
\vsize 25.0truecm
\topmargin=0cm
\topskip=0cm

\begin{titlepage}

\rightline{\bf hep-lat/9406015}
\rightline{UvA-ITFA 94-17}
\rightline{UCSD/PTH 94-07}
\vskip 3mm
\rightline{June 1994}

\baselineskip=20pt plus 1pt
\vskip 0.5cm

\centerline{\LARGE Fermion production despite}
\centerline{\LARGE fermion number conservation}
\vskip 0.5cm

\centerline{\bf Wolfgang Bock}
\centerline{\sf bock@sdphjk.ucsd.edu}
\medskip
\baselineskip=14pt
\centerline{\it Department of Physics,}%
\centerline{\it University of California, San Diego}
\centerline{\it Gilman Drive 0319, La Jolla, CA 92093-0319}
\centerline{\it USA}
\medskip

\centerline{
{\bf James E. Hetrick\footnote{ Address after 1 Oct. 1994: Physics
Dept., University of Arizona, Tucson AZ 85721, USA}
}~~~~~\raisebox{-1.0ex}{and}~~~~~~~~{\bf Jan Smit}}
\centerline{
{\sf \ ~~~~~hetrick@phys.uva.nl}~~~~~~~~~~~~~~~~{\sf
jsmit@phys.uva.nl}}
\medskip
\centerline{\it Institute for Theoretical Physics}
\centerline{\it University of Amsterdam}
\centerline{\it Valckenierstraat 65, 1018-XE Amsterdam}
\centerline{\it The Netherlands}

\vskip 0.7cm
\baselineskip=12pt plus 1pt
\parindent 20pt
\centerline{\bf Abstract}
\textwidth=6.0truecm
\medskip

\frenchspacing

Lattice proposals for a nonperturbative formulation of the Standard
Model easily lead to a global U(1) symmetry corresponding to exactly
conserved fermion number. The absence of an anomaly in the fermion
current would then appear to inhibit anomalous processes, such as
electroweak baryogenesis  in the  early universe. One way to
circumvent this problem is to formulate the theory such that this U(1)
symmetry is explicitly broken. However we argue that in the framework
of spectral flow, fermion creation and annihilation still in fact
occurs, despite the exact fermion number conservation.  The crucial
observation is that fermions are excitations relative to the vacuum,
at the surface of the Dirac sea.  The exact global U(1) symmetry
prohibits a state from changing its fermion number during time
evolution, however nothing prevents the fermionic ground state from
doing so.  We illustrate our reasoning with a model in two dimensions
which has axial-vector couplings, first using a sharp momentum cutoff,
then using the lattice regulator with staggered fermions.  The
difference in fermion number between the time evolved state and the
ground state is indeed in agreement with the anomaly. A study of the
vacuum energy shows that the perturbative counterterm needed for
restoration of gauge invariance is insufficient in a nonperturbative
setting.  For reference we also study a closely related model with
vector couplings, the Schwinger model, and we examine the emergence
of the $\theta$-vacuum structure of both theories.
\nonfrenchspacing

\end{titlepage}

\textheight=20.5cm
\headsep=2.0cm
\vsize 21truecm

\section{Introduction}

A tractable nonperturbative formalism for the analysis of chiral
gauge theories is an outstanding problem in modern field theory. Among
the difficulties it has been argued that lattice formulations (for
reviews see \cc{REVIEW,Sm88}) or indeed any regularization, cannot
yield the correct physics if it implies an exact global symmetry which
should be broken by anomalies \cc{Sm88,EiPr86,Ba91}.  For example,
such models would be unable to describe anomalous fermion number
violation in the Standard Model, which for example, plays an important
role in current explanations of the observed baryon asymmetry in the
universe.

A lattice regularized version of the Standard Model is of course
supposed to describe these effects correctly. However typical forms of
the lattice fermion action, $S_f = -\sum_{xy} \psb_x D_{xy}
\ps_y$ have an exact global U(1) invariance $\ps\ra\exp(i\om)\ps$,
$\psb\ra\exp(-i\om)\psb$ and since the lattice fermion measure is
generally also invariant, one expects this global symmetry to be
accompanied by exact fermion number conservation, suggesting that such
models cannot display the anomalous fermion production of the Standard
Model. We shall call this the electroweak U(1) problem.

Some recent discussions have focused on the question of heavy fermion
decoupling \cite{REVIEW,Ba91,BaNa92,DuMa91}, but the U(1) problem is
more general. We also remark that its resolution should not depend on
alternative definitions of the fermion number current, constructed
such that their divergence has the right anomaly (see
e.g. \cc{DuMa91,BoSm94a}). Once the action and measure in the path
integral are defined nonperturbatively, the stage is set, and we
should just have to see what physics emerges.

One way out would be to construct lattice actions such that they have
no unwanted exact global symmetries
\cite{EiPr86,Ba91,Pr91,MaRoTe92,Sm92a}
and it is reasonable to assume that the violation of the global
symmetry turns in the scaling region into the desired anomaly
\cite{KaSm81}.

Here we want to reconsider this standard lore and study the effects of
the unwanted global U(1) symmetry in general terms. We examine the
spectral flow to see whether creation or annihilation of fermions,
i.e. fermionic particles, takes place in a simple model in 1+1
dimensions which has the exact global U(1) symmetry corresponding to
fermion number conservation. We study the spectrum of the fermionic
Hamiltonian in an external gauge field which changes slowly in time,
the idea being that we consider time slices of an instanton-like
configuration, which starts as a vacuum gauge field, goes through a
sphaleron configuration and ends up as a different vacuum gauge field
related to the first by a topologically non-trivial (`large') gauge
transformation. If the change with time is very slow, we can make use
of the adiabatic theorem \cc{OLD} and deduce the evolution of the
states by continuity, following the eigenstates of the time dependent
Hamiltonian. We can then check explicitly whether an initial fermionic
vacuum state has evolved into a state with particles produced by the
gauge field in accordance with the anomaly. Particle annihilation
proceeds of course by the inverse process.  For a recent discussion of
anomalies and spectral flow see ref.~\cc{NN}; an earlier spectral flow
analysis in lattice gauge theory was given in ref.~\cc{AmGr83} and
more recently in the context of domain wall fermions \cite{Ka92} in
ref.~\cite{CrHo94}.

Based on the spectral flow analysis we argue that fermion production
occurs {\em despite} the exact fermion number conservation. The
crucial observation is that fermions are excitations relative to the
vacuum, at the surface of the Dirac sea, which are defined as the
ground states in external fields that are `pure gauge'. The exact
global U(1) symmetry prohibits a state from changing its fermion
number during time evolution, however nothing prevents the fermionic
ground state, the state of lowest energy which depends on the
adiabatic external gauge field, from changing {\em its} fermion
number. We show that the difference in fermion number between the time
evolved state and the ground state after the sphaleron transition is
indeed in agreement with the anomaly.

Our arguments do not really depend on lattice regularization.  Still,
we would like to indicate how they might fit into a possible lattice
formulation of chiral gauge theories, taking into account the
experience gained from investigations over the past years showing the
failure of many promising proposals \cite{REVIEW}.  Recall an often
followed practice in the continuum: the fermions are integrated out
first and from the resulting determinant a well defined effective
action is constructed for the bosonic fields.  This can be done in a
variety of ways; in general it requires the addition of counterterms
to get a finite answer upon removing the fermionic regulator, assuming
the fermions are in an anomaly free representation of the gauge
group. Subsequently, the integration over the bosonic fields is to be
performed. One first removes the fermion regulator and then the boson
regulator. We see no reason in principle why such a practise cannot be
followed on the lattice.  It means dealing with lattice fermions in
smooth external gauge fields, a relatively simple situation for which
any lattice fermion method should work. Counterterms are usually
needed to restore (chiral) gauge invariance, and with a gauge
invariant effective action the introduction of ghost fields
\cc{BoMa89} may be avoided as in non-chiral lattice gauge theory (the
problem noted in \cite{RoSa93} need not apply).  Taking the fermion
lattice distance all the way to zero first (`the desperate's method'
\cite{Sm88,GoSc92}), may not be necessary if the violation of
gauge invariance can be controlled so that gauge symmetry restoration
may be invoked \cite{BoSm93b}. Obviously, much work still remains to
be done in this direction, and our present study of the subtleties of
anomalous fermion production seems a necessary step, which is also of
interest in its own right. A preliminary account of this work has
already been presented in ref.~\cc{BoHe94}.

The outline of the paper is as follows: In Sect.~2 we introduce the
two-dimensional massless models, the Schwinger model and its
equivalent axial version (axial QED$_2$), which we shall use in this
paper to illustrate our reasoning.  In Sect.~3 we first discuss the
spectral flow of these models with a sharp momentum cutoff on the
number of modes. We argue that fermion creation and annihilation in
axial QED$_2$ is possible in spite of the exact U(1) symmetry. The
sharp momentum cutoff is unsatisfactory because it is both non-local
and gauge variant. Local lattice versions of the two models are
introduced in Sect.~4, using staggered fermions. The lattice axial
QED$_2$ model lacks gauge invariance, which can be restored for smooth
external gauge fields with a mass counterterm for the gauge field. We
then study the spectral flow in both models to see how the properties
of the well understood vector model compare with those of the axial
model. Fermion creation and annihilation takes place in the same way
as in the toy models with the sharp momentum cutoff. There are
furthermore interesting aspects to the way the counterterm restores
gauge invariance to the ground state energy in the axial model, and to
the way the Dirac sea acquires a gauge invariant bottom in the vector
model. We summarize our conclusions in sect.~5.  For clarity of
presentation we have delegated many details of the staggered fermion
formalism and the calculation of the energy spectrum to appendices
A--C.

\section{Axial and vector QED$_2$ in the continuum}
The massless axial QED$_2$ model is given in the continuum by the
following (real time) action,
\bea
S\!\!&=&\!\! -\int d^2 x \frac{1}{4e^2} F_{\mu\nu}F^{\mu\nu} + S_f
\;, \label{FSQ}\\
S_f\!\! &=&\!\! -\int d^2 x \; \jb \g^{\mu}
(\dmu + iA_{\mu}\g_5)\j, \lb{AXQED}
\eea
where $\g^1=\g_1=\s_1$, $\g^0 = -\g_0 = -i\s_2$, $\g_5 = \s_3$.
We take $A_{\mu}$ to be an external gauge field; only the fermion
fields are quantized. The model is invariant under local axial gauge
transformations: $\j(x) \ra \exp( i \om(x) \gm_5)
\j(x)$, $\jb(x)\ra \jb(x) \exp( i \om(x) \gm_5)$, $A_{\m}(x)
\ra A_{\m}(x) + \dmu \om(x)$, with gauge current $j^{\mu}_5 = i \jb
\g^{\mu} \g_5 \j$.
The action is furthermore invariant under the global U(1) symmetry
$\j(x) \ra \exp(i \al) \j(x)$, $\jb(x) \ra \jb(x) \exp(-i
\al)$, with a corresponding vector current
$j^{\mu} = i \jb \g^{\mu}\j$.  According to standard lore the gauge
current, upon quantization, has to be divergence free, $\dmu
j^{\mu}_5=0$, while the divergence of the vector current becomes
anomalous
\be
\dmu j^{\mu} = - 2 q  \;,\;\;\;\;
q =\frac{1}{4\pi} \e^{\m \n} F_{\m \n} \equiv \dmu C^{\m} \;,
\label{DIV}
\ee
where $C^{\m}(x)=(1/2\pi) \e^{\m \n} A_{\n}(x)$ is the Chern-Simons
current and $q(x)$ is the topological charge density ($\e_{01} = +1$);
the topological charge is defined by
$\int d^2 x \;q(x)$.
Fermion and Chern-Simons numbers are given by the spatial
integrals over the time components of the corresponding currents,
$Q=\int dx^1\; j^0 (x)$, $C=\int dx^1\; C^0 (x)$. For later purposes
we also introduce here the axial charge $Q_5=\int dx^1\; j^0_5(x)$
which is exactly conserved in this model. From eq.~(\eq{DIV}) one can
deduce the equation
\be
Q(t)-Q(0) = - 2 ( C(t) - C(0) ) = -2 \int_0^t dt\,\int dx\, q
\;,
  \;\;\; x\equiv x^1,\;\;t \equiv x^0\;,
\lb{DF}
\ee
which relates a change in the Chern-Simons number to a change in
fermion number.  The aim of this paper is to investigate whether, in
those models which have the exact global U(1) symmetry and where $Q$
is thus conserved, a change in $C$ due to a sphaleron transition can
still give rise to a change in the number of fermions in the
spirit of eq.~(\eq{DF}).

By charge conjugation of the right-handed fermion fields,
\be
\j_R =(\jbp_R {\cal C})^T\;, \;\;\;
\jb_R = - ({\cal C}^{\dagger} \jp_R)^T\;, \;\;\;
\j_L =\jbp_L \;, \;\;\;
\jb_L = \jp_L,  \lb{CG}
\ee
with ${\cal C}$ the charge conjugation matrix, or alternatively to
only the left-handed fields, one can show that axial QED$_2$ is
equivalent to the massless Schwinger model (vector QED$_2$),
\be
S_f^\prime = -\int d^2 x \jbp \g^{\mu}(\dmu - i
A_{\mu})\jp\;. \lb{VCQED}
\ee
For the moment we will label the quantities in the vector model with a
prime for clarity.  The local axial gauge symmetry of the action
(\eq{AXQED}) has turned, after the transformation (\eq{CG}), into a
local vector gauge symmetry for the $\jp$ fields, $\jp(x)
\ra \exp(i \om(x)) \jp(x)$, $\jbp(x) \ra \jbp(x) \exp(-i
\om(x))$, and vice versa the global vector invariance into a global
axial invariance $\jp(x) \ra \exp(i \al \gm_5 ) \jp(x)$, $\jbp(x) \ra
\jbp(x) \exp(i \al \g_5)$.  The corresponding currents and charges
transform under (\eq{CG}) into each other,
\bea
&& j_5^{\mu}= i \jb \g^{\mu} \g_5 \j = -i \jbp \g^{\mu} \jp =
-j'^{\mu} \lb{CAV} \nnn && j^{\mu}= i \jb \g^{\mu} \j = -i \jbp
\g^{\mu} \g_5 \jp = -j_5^{\prime \mu} \lb{CVC} \nnn && Q=-Q_5^{\prime}
\;,\;\;\;\;Q_5=-Q^{\prime} \;. \lb{FF}
\eea
We notice that a mass or Yukawa term added to the actions (\eq{AXQED})
and (\eq{VCQED}) would transform under (\eq{CG}) into a corresponding
Majorana mass or Yukawa term.  In this paper we will not include these
terms. For a discussion of the spectral flow in QED in presence of a
bare mass term see ref.~\cc{AmGr83}.

In the following we use a finite spatial extent of length $L$, with
periodic boundary conditions for the gauge field and antiperiodic
boundary conditions for the fermion fields. We consider gauge fields
with vanishing time component, $A_0=0$, and choose the Coulomb gauge
specified by $A_1(x,t)=A(t)$; then the time dependence is expressed by
$A$-dependence. In this gauge the Chern-Simons number $C=-AL/2\pi$,
and going through an instanton-like configuration means that $C$
changes by one unit.  Values $A= 2\pi k/L$ with integer $k$ are `pure
gauge', since for these values we can write $A=\Om i\partial_x\Om^*$,
with $\Om=\exp(i2\pi k x/L)$ a periodic gauge transformation with
winding number $k$. When $A= 2\pi k/L$ the Chern-Simons number takes
integer values $C=-k$.  An example of such gauge fields is given by
the configuration with constant electric field $f=F_{01}$,
\be
A_0 = 0,\;\;\; A_1(x,t) = A(t) = -f \; t. \label{CFS}
\ee
For $f=2\pi/LT$ we have $C(0) = 0$, $C(T)=1$, and the adiabatic limit
corresponds to $T\ra\infty$.

The ground state $|0, A\rag$ of the fermionic Hamiltonian $H(A)$ is by
definition the state with lowest energy. When $A$ is pure gauge, these
ground states are to be identified with the vacuum.
In Hilbert space,
gauge transformations on $A$ induce unitary transformations, and we
have to identify states related by gauge transformations, hence also
all vacua with differing integer Chern-Simons numbers.

A quantity which we will use in our analysis is the ground state
energy $E_0(A)$, $H(A)|0, A\rag =E_0(A)|0, A\rag$. It can be obtained
from the partition function at inverse temperature $\bt$ in the limit
$\bt\ra\infty$
\be
E_0(A)= -\frac{1}{\bt}\ln \Tr e^{-\bt H(A)}, \;\;\;
\bt\ra\infty\;,
\ee
or in terms of the euclidean effective action in a space-time volume
$L\times \bt$,
\bea
S_{\rm eff}(A) &=& \ln \int D\jb D\j\, \exp(S_f^{\rm eucl}), \lb{Seff}
\\ &\ra& -\bt E_0(A),\;\;\; \bt\ra\infty.
\label{BINFTY}
\eea
We shall use the euclidean formalism for our lattice models.  The
quantity $E_0(A)$ can be evaluated as (see for example \cite{HH}
and references therein)
\be
E_0(A) =
\frac{2}{\pi L} \sum_{n=1}^\infty \frac{(-1)^n}{n^2}\cos(n LA)
+ {\rm const}.
\ee
The sum is clearly periodic in $A$ with period $2\pi/L$ and evaluates,
in each period, to a quadratic potential.
\be
E_0(A)- E_0(0) = \frac{L}{2 \pi} A^2,\;\;\; -\frac{\pi}{L} < A <
+\frac{\pi}{L}, \;\;\; {\rm mod}\,(2\pi/L).  \label{E0}
\ee
{}From (\ref{E0}) we can read off the dynamically generated mass of
the Schwinger Model, $m^2 = e^2/\pi$ (the factor $e^2$ is due to
the $F^2$ term in (\ref{FSQ})). For a treatment of the Schwinger model on a
circle in the hamiltonian formalism, see refs. \cite{Ma85}.
\section{A model with a sharp momentum cutoff}
Before giving a full treatment using lattice fermions we first
consider a model of fermions regulated by a sharp cut-off in momenta
$\Lm$.  Such a regulator captures essential features we wish to study
in lattice regularization, namely, exact global invariance and
restriction to a finite number of modes for the $\psi$ field. It is
not gauge invariant, which it shares with the class of chiral lattice
models we have in mind, and moreover it is non-local, whereas the
lattice models are local. The locality of the lattice models allows
for restoration of gauge invariance by a local mass counterterm
$\propto \int d^2 x\, A_{\mu} A^{\mu}$.

{}From the above actions (\eq{AXQED}) and (\eq{VCQED}), the axial
QED$_2$ hamiltonian $H$ and the equivalent vector QED$_2$ hamiltonian
$H'$ are given by
\bea
H\!\!&=&\!\!\frac{1}{L}\sum_p[\psd_R(p)\ps^{\mbox{}}_R(p)(p+A) +
\psd_L(p)\ps^{\mbox{}}_L(p)(-p+A)] -AN/2\,,\label{AXCUT} \\
H'\!\!&=&\!\!\frac{1}{L}\sum_p[\psdp_R(p)\ps'_R(p)(p-A) +
\psdp_L(p)\ps'_L(p)(-p+A)],\label{VECCUT}
\eea
where the summation is over momenta $p = (n-1/2)2\pi/L \in
[-\Lm,\Lm]$ (antiperiodic boundary conditions for the fermions),
$N$ is the total number of modes, and $L$ and $R$ denote the left-and
right-handed projections $\gm_5 \ra \mp 1$.  The Fourier modes $\ps(p)
= \int_0^L dx\exp(-ipx)\ps(x)$ have the commutation relations
$\{\j(p),\j^{\dagger}(q)\} = L \dl_{p,q}$.  In axial QED$_2$ the
spatial component of the gauge field couples to fermion number
\bea
Q&=&\int dx\, \half[\j^{\dagger}(x),\j(x)]\nonumber\\ &=& \frac{1}{L}
\sum_p [\psd_R(p)\ps^{\mbox{}}_R(p) + \psd_L(p)\ps^{\mbox{}}_L(p)] -
N/2. \label{QFORM}
\eea
Note that upon quantization we have anti-symmetrized the field
operators in $Q$, which is necessary for it to change sign under
charge conjugation.  Charge conjugation (\ref{CG}) on the right handed
fields only, relating the axial and vector model, is given by $\ps'_R
= \psd_R$, $\psdp_R = \ps_R$. Denoting this conjugation by ${\cal
C}_R$, we have $H'={\cal C}_R H {\cal C}_R^{\dagger}$, so that $H'$
has the same spectrum as $H$.

In these models regularized by a sharp momentum cutoff, there are {\em
two} conserved charges: both $Q$ and
\be
Q_5 = \int dx\, \j^{\dagger}(x)\g_5\j(x) =\frac{1}{L}
\sum_p [\psd_R(p)\ps^{\mbox{}}_R(p) - \psd_L(p)\ps^{\mbox{}}_L(p)]
\ee
commute with $H$. In the vector version the two charges are given by
$Q' = -Q_5$, $Q'_5 = -Q$ (cf. eq. \ref{FF}).

The spectral flow of the eigenmode energies $\e$ in the axial model is
very simple. Disregarding the mode-independent term $-AN/2$ we have
\be
\e_{\c}(p,A) = \c p + A \;,  \lb{AXVECE}
\ee
where the chirality $\c = \pm 1$ is the eigenvalue of $\g_5$
($\c=+1$ ($-1$) for the $R$ ($L$) modes). Each
mode's energy increases linearly with $A$.  In the vector version
\be
\e^\prime_{\c}(p,A) = \c (p - A)\;, \lb{VECE}
\ee
and the $R$ ($L$) modes decrease (increase) linearly with $A$ as shown
in fig. 1.
%
%
\begin{figure}[htb]
   \epsfxsize=14.5cm \centerline{\epsffile[37 225 525
545]{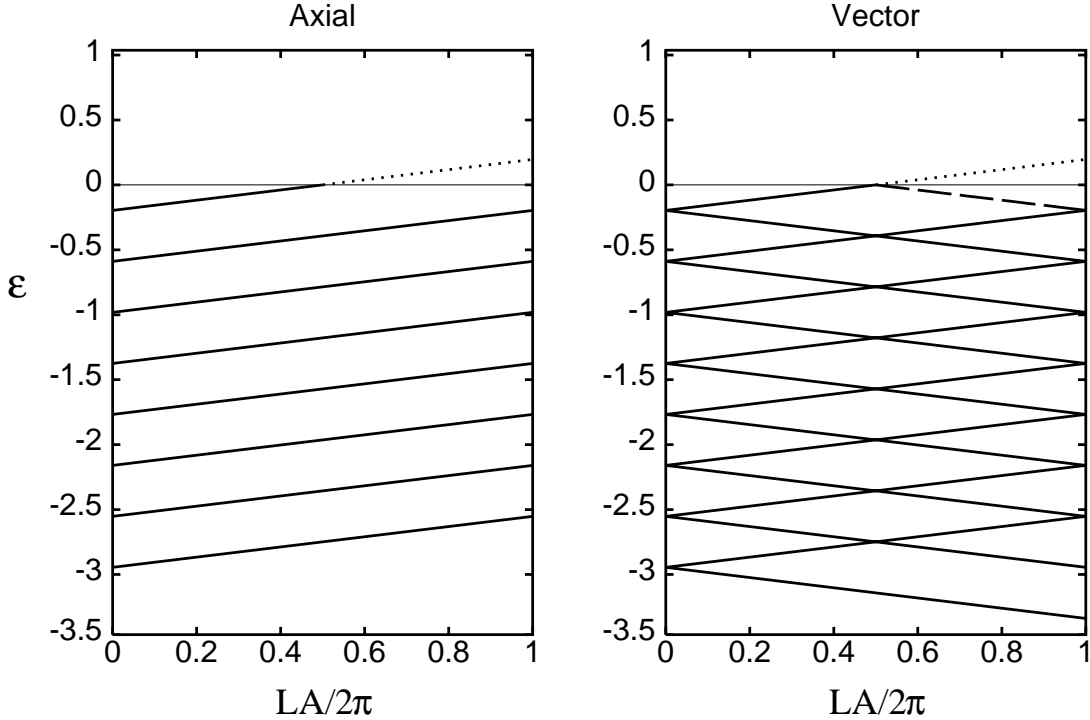}}
\vspace{1cm}
\caption{\em Occupation of the ground state $|0,A\rag$  (solid and
         dashed) and the adiabatically time evolved state
$|\Psi,A\rag$ (solid and dots) in the momentum cutoff models.}
\vspace{0.5cm}
\end{figure}

Consider now the flow of the state $|\Ps,A\rag$, which starts out as
the vacuum state at $A=0$, i.e. the Dirac sea with all negative energy
modes occupied. Each mode is doubly degenerate ($L$ and $R$). In the
axial model the $\e_p$ flow upwards with $A$ and at $A=2\pi/L$ they
have taken the place of their predecessor, except for the two modes
starting out as $\e=-\pi/L$ and ending up at $\e=+\pi/L$.  The quantum
numbers $Q$ and $Q_5$ are conserved, both are zero for all states
$|\Ps,A\rag$, as for the initial vacuum state $|0,0\rag$.  The energy
$E_{\Ps}$ of the state $|\Ps, A\rag$ also happens to be constant in
this regularization: each mode shifts upwards by $A$ and the $N/2$
modes contribute $AN/2$, which is compensated by the mode-independent
term $-AN/2$. However, the ground state $|0,A\rag$ looses two occupied
modes half way at $A=\pi/L$, when $\e = -\pi/L+A$ crosses zero. The
final vacuum energy $E_0$ at $A=2\pi/L$ differs from the initial one,
due to the lack of gauge invariance of the sharp momentum cutoff.

The important point is that the final ground state has two occupied
states ($L$ and $R$) less than the initial ground state, so the final
vacuum quantum numbers are $Q=-2$ and $Q_5=0$. Hence,
\be
\Delta Q \equiv
(Q_{\Ps}-Q_{0})_{\rm final} - (Q_{\Ps}-Q_{0})_{\rm initial}
=+2\,,
\ee
just as expected from the anomaly in the vector current, $\Delta Q =
-2\Delta C$.  In the vector model the $L$ modes move upwards and the
$R$ modes move downwards, with $(Q_{5,\Psi})_{\rm initial} =
(Q_{5,\Psi})_{\rm final} = 0$. Midway, the ground state looses an $L$
mode and gains an $R$ mode, such that $(Q_{5,0})_{\rm initial} =0$,
$(Q_{5,0})_{\rm final} = +2$ and $\Delta Q'=0$, $\Delta Q'_5=-2
=2\Delta C$.  In the axial model we have the creation of two particles
($L$ and $R$), whereas in the vector model we have the creation of a
particle ($L$) and an antiparticle ($R$) (an $R$-hole).

In both models the final energy difference of $|\Ps,A\rag$ with
$|0,A\rag$ is two
units $\pi/L$, which corresponds to the creation of the two particles
(using the generic term `particle' also for the antiparticles).  It is
amusing that this can be interpreted in classical terms, as the work
which is provided by the external field in creating two fermions
\cc{NN}. The time component of the Lorentz force
$d p_{\mu}/dt = \pm F_{\mu\nu}
dx^{\nu}/dt$ on a particle of charge $\pm 1$ can be integrated for the
gauge field of constant field strength (\ref{CFS}).  Using $dx^1/dt =
\pm 1$ for the massless particles, and taking into account that they
appear midway at $t=T/2$, this gives a final energy per particle
$p^0(T) = \int_{T/2}^T dt\, dp^0/dt = \pi/L$.

Obviously, the ground state energy in these cutoff models is very
different from the continuum form (\ref{E0}).  It is given by
$E_0(A)-E_0(0) =0$ for $0\leq A \leq \pi/L$ and $E_0(A)-E_0(0) =
-2(A-\pi/L)$ for $\pi/L \leq A \leq 2\pi/L$.  A `counterterm'
piece-wise linear in $A$ would be required to rectify the
situation. This is at variance with the expectation that a simple
quadratic mass counterterm $\propto
\int d^2 x\, A_{\mu}A^{\mu}$ should be sufficient to restore gauge
invariance. The sharp momentum cutoff is also unsatisfactory because
it is non-local. The local lattice models
to which we shall turn in the next section
are well behaved in this
respect.
\section{Level shifting in the lattice models with staggered fermions}
Our lattice axial and vector QED$_2$ models are formulated in terms of
staggered fermion fields. In order to keep the technicalities to a
minimum, we have collected the details of this formulation to the
appendices and here we just record the essentials. We use the
Euclidean lattice formulation with the partition functions given by
\be
Z = \int (\prod_x d\cb_x d\c_x)\, \exp S_f, \lb{PINT}
\ee
with
\be
S_f = - \sum_{x\m} \h_{\m x} \half (\cb_x U_{\mu x} \c_{x+\hmu}-
\cb_{x+\hmu} U_{\mu x}^* \c_{x})\;,
\lb{CHIQED}
\ee
for vector QED$_2$ and
\bea
S_f\!\!  &=&\!\!  -\sum_{x} \left\{ \sum_{\mu}\cos A_{\mu x}\,
\et_{\mu x} \half (\cb_x\c_{x+\hm} -
\cb_{x+\hm}\c_x)\right. \nonumber\\
&&\left.\mbox{} + \sum_{\mu\nu}\sin A_{\mu x}\,\ep_{\mu\nu}
\et_{\nu x} \half (\cb_x\c_{x+\hat{\nu}}
+ \cb_{x+\hat{\nu}}\c_x) \right\}
+ \sum_{x\mu} \t\half A_{\mu x}^2\;,
\lb{CHIAX}
\eea
for axial QED$_2$.  We use lattice units, in which the lattice
distance $a=1$.  $U_{\mu x} = \exp(-iA_{\mu x})$ is the lattice gauge
field, the $\et_{\m x}$ are the usual staggered fermion sign factors
representing the gamma matrices, $\et_{1x}=1$, $\et_{2x}=(-1)^{x_1}$
($\g_2 = i\g^0$, $\e_{12}=+1$). The last term in (\ref{CHIAX}) is a
counterterm. The vector model is gauge invariant; the axial model is
not, however invariance is restored in the scaling region by the
addition of this counterterm.
The derivation of the continuum models (\eq{VCQED}) and (\eq{AXQED})
from the staggered fermion actions (\eq{CHIQED}) and (\eq{CHIAX}) is
sketched in appendix A, where we give also the perturbative result for
the coefficient $\tau$ of the counterterm. A nonperturbative extension
of the counterterm is discussed below.

The exact U(1) phase invariance of the actions and measure in the path
integrals above is obvious. For the vector model this is just the
global limit of its local gauge invariance.  For the axial model we
have an exact global $U(1)$ symmetry leading to fermion number
conservation.  To derive the divergence relation of the current
associated with this symmetry we replace $\c_x \ra \exp(i\o_x) \c_x$
and $\cb_x \ra \exp(-i\o_x) \cb_x$ in the path integral (\eq{PINT})
and note that it is invariant under this transformation. After
collecting terms linear in $\o_x$ we find from the fact that their
coefficients must vanish, the exact current divergence relation on the
lattice,
\be
\sum_\m \lag j_{\m x}-j_{\m x-\hmu} \rag_{\c} =0 \;,
\ee
where the brackets denote the average over the fermion fields.
For the vector theory the current is given by
\be
j_{\mu x} = i\et_{\mu x}\half (\cb_x U_{\mu
x}\c_{x+\hm}+\cb_{x+\hm}U_{\mu x}^*\c_x)\;,
\label{JVEC}
\ee
while for the axial theory
\bea
j_{\mu x} &=& i \cos A_{\mu x}\, \et_{\mu x} \half (\cb_x
\c_{x+\hm}+\cb_{x+\hm}\c_x) \nonumber \\
& &~~~~~~~~+i\sin A_{\nu x}\, \ep_{\mu\nu} \et_{\mu x}
\half (\cb_x \c_{x+\hat{\mu}} - \cb_{x+\hat{\mu}}\c_x)
\;.\label{JAX}
\eea
The current in the vector theory (\ref{JVEC}) is of course gauge
invariant, but this is lacking in the axial theory (\eq{JAX}).
Presumably the combination $j_{\mu x} +2iC_{\mu x} = j_{\mu x} +
(i/\pi) \ep_{\mu\nu}A_{\nu}$, which has the correct euclidean anomaly,
will become gauge invariant in the scaling region and can be compared
with the continuum current in (\eq{DIV}).

The above staggered fermion models actually describe axial or vector
QED$_2$ with two flavors.  The corresponding SU(2)$_L\times$SU(2)$_R$
flavor symmetry is reduced on the lattice to the exact global U(1)
symmetry $\c_x\ra\exp(i\o\e_x)\c_x$, $\cb_x\ra\exp(i\o\e_x)\cb_x$,
$\e_x=(-1)^{x_1+x_2}$, which corresponds to a flavor non-singlet axial
transformation in the scaling region
\cite{DoSm83}.  The charge corresponding to this
symmetry is given in appendix B. As long as the gauge field is
external, the two flavors are an inessential extension of the previous
models.

Other currents may be constructed which reduce in the scaling region
to the expected continuum currents.  These are somewhat arbitrary
because they do not correspond to symmetries of the action, although
natural choices exist \cite{BoSm94a}.  It has been shown in
ref. \cite{ShTh82} that the natural choice for the axial current in
the gauge invariant vector QED$_2$, for example, has the expected
anomalous divergence equation; see \cite{SmVi88} for $U(4)\times U(4)$
currents in staggered fermion QCD.

The gauge potential $A_{\mu x}$ is taken to be independent of
euclidean (`imaginary') time $x_2$. It depends parametrically on the
real time $t$. From the euclidean path integrals above we can derive
the transfer operator $T$ in the standard way (cf.  appendices B and
C), $Z=\Tr T^{\bt/2}$, where $\bt/2$ is the number of pairs of time
slices and $\bt$ the inverse temperature.  The hamiltonian $H$ is
defined in terms of $T$ by $T=\exp(-2H)$, such that $Z=\Tr \exp(-\bt
H)$.

For the simple case $A_1 = A(t)$ (i.e. independent of $x_1$ and
$x_2$), $A_2=0$,
we can obtain the spectrum of the transfer operator explicitly, and
from this the energy levels of the associated hamiltonian as a
function of $t$ --- the spectral flow. From the calculations outlined
in appendices A-C we find the following formulas for the energies. For
vector QED$_2$:
\bea
\e_{\c}(p,A) &=& \e_{0\c}(p-A)\,,\nonumber\\
\e_{0\c}(p) &=& \ln(\c\sin p + \sqrt{\sin^2 p + 1})\,,\;\;\;
{\rm or}\;\; \sinh \e_{0\c} = \c\sin p,\label{EEQED}\\ p&=&
n\frac{2\pi}{L}-\frac{\pi}{L}\,,\;\;\; n = -
\frac{L}{4}+1,\cdots,\frac{L}{4}. \nonumber
\eea
Here the $\e_{0\c}(p)$ are the free staggered fermion energies and
$\c=\pm 1$ is the chirality of the mode, the eigenvalue of $\g_5$.
Each mode is doubly degenerate because of the two flavors. The
eigenvalues of the hamiltonian,
\be
E = L\ln 2 +
\kern-0.5em \sum_{\stackrel{\scriptstyle{\rm occupied}}{\rm modes}}
\kern-0.5em\e_{\c}(p,A),
\ee
consist of the sum of the occupied mode energies plus an overall
constant $L\ln 2$, which could be avoided by redefining the fermion
measure in the partition function.
{}For axial QED$_2$:
\be
\e_{\c}(p,A) = \e_{0\c}(p) + \ln\sqrt{\frac{1+\sin A}{1-\sin A}} \; ,
\label{EEAX}
\ee
where we assume $\cos A > 0$ (below we shall furthermore restrict
ourselves to $|A|<\pi/4$). In this case
\be
E = L\ln 2 - L\ln(1+\sin A) +
\kern-0.5em \sum_{\stackrel{\scriptstyle{\rm occupied}}{\rm modes}}
\kern-0.5em \e_{\c}(p,A),
\ee
where $E$ does not include the counterterm. The term $-L\ln(1+\sin A)$
is the analogue of the `charge term' $-AN/2$ in (\ref{AXCUT}),
generalized to two flavors (in lattice units $L=Na=N$).

\begin{figure}[htb]
\vspace{-1cm}
   \epsfxsize=14.5cm
   \centerline{\epsffile[37 225 525 545]{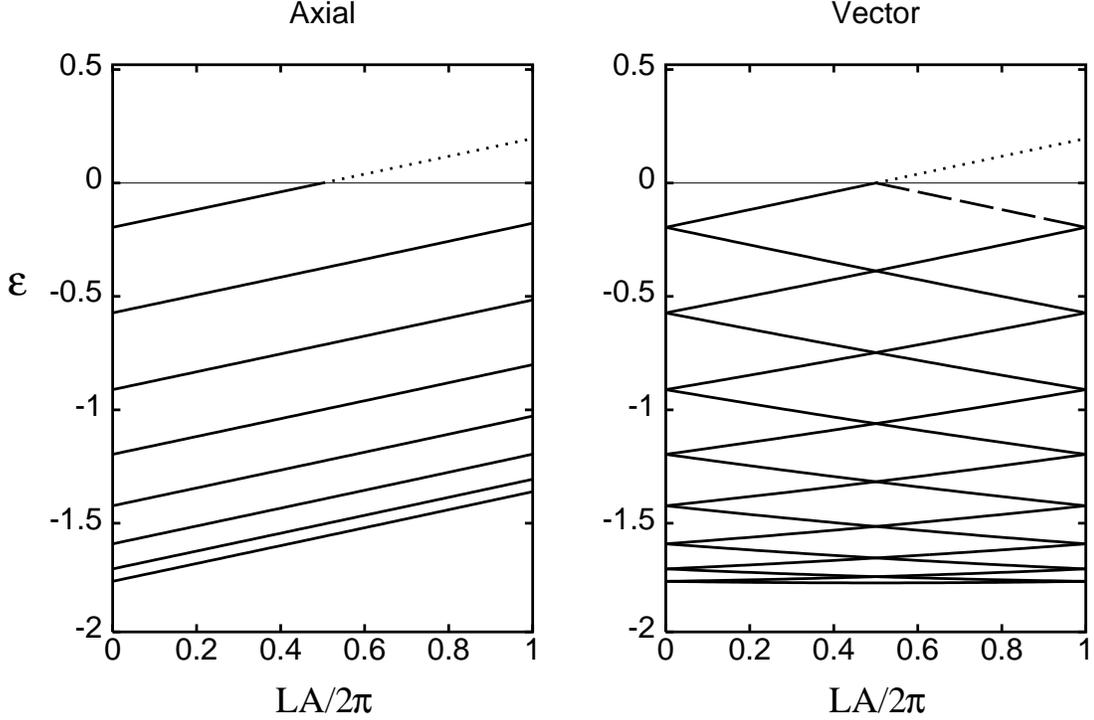}}
\caption{{\em  Spectral flow for the axial and vector models
with staggered fermions. The energies $\e_{\c}(p,A)$ are shown as
function of
$LA/2\pi$ for $L=32$. Ground state
$|0,A\rag$ and $|\Psi, A\rag$ are represented as in fig. 1}}
\vspace{0.5cm}
\label{FLOW}
\end{figure}

We have plotted in fig.~\ref{FLOW} the energies $\e_{\c}(p,A)$ as a
function of $AL/2\pi$ in the interval between $0$ and $1$, for a
lattice of size $L = 32$.  Each energy eigenvalue in (\eq{EEQED}) and
(\eq{EEAX}) appears twice because of the two flavors. For momenta $p$
and gauge fields $A$ which are small compared to the cutoff $\pi/2$,
(\eq{EEQED}) and (\eq{EEAX}) reduce to the linearly spaced modes in
formulas (\eq{VECE}) and (\eq{AXVECE}). The spectra close to the
surface of the Dirac sea, for small $A$, are as in the continuum. For
the modes near the maximal negative or positive energies we clearly
see regularization effects.  But most importantly, the interpretation
of anomalous fermion creation and annihilation for the spectra in
fig.~\ref{FLOW} is just as in the previous section: a given quantum
state respects the global $U(1)$ invariance of the action, while the
vacuum adjusts, producing or absorbing the charge in concordance with
the anomaly.

There are furthermore some interesting details which are worth
mentioning. In the vector case the two spectra describing the vacuum
at $A=0$ and $A=2\pi/L$ are identical, in accordance with gauge
invariance. As in the sharp cut-off model, the $L$-modes move
upwards, whereas the $R$-modes move downwards. This however does not
hold for the upper- and lower-most modes. For example, the maximal
negative energy mode appears to start out right handed
$A=0$ (going down) while ending up left handed
(going up). This effective chirality flip occurs because the
staggered fermion analogue of $\g_5 (p-A)$ is $\g_5\sin(p-A)$ the
slope of which, $(\partial/\partial A) \, \g_5\sin(p-A) =
-\g_5\cos(p-A)$, changes sign at $A=\pi/L$ when the momentum is
minimal, $p=-\pi/2 + \pi/L$ (at the bottom of the Dirac sea).
{}From a more general point of view we expect in coupled
staggered fermion theories to be able to recover $\gm_5$ only in
the scaling region, i.e. for sufficiently small $p$.
Far from the scaling regime the labels `$L$' and `$R$' may lose
their usual meaning. In the present situation, however, the
perfectly smooth (constant) $A$ allows for a more detailed
identification of $\gm_5$ also outside the scaling region, which
leads to the chirality flip in the minimum energy modes at
$A=\pi/L$.

The chirality of an eigenmode may be viewed as the eigenvalue of an
axial charge $\widetilde{Q_5}$, as defined in (\ref{Qtilde}), which
commutes with the transfer operator. This $\widetilde{Q_5}$ is the
analogue of $\int dx\,
\j^{\dagger}\gm_5\j - 2C$
in continuum considerations and it is
not gauge invariant. On the other hand the natural gauge
invariant but not conserved $Q_5$ constructed from the current
$j_{\mu\, 5} = -i\ep_{\mu\nu} j_{\nu}$ gives only sensible
chiralities in the scaling region (cf. (\ref{JVEC}), (\ref{Q5})).

The behavior under large gauge transformations is also
interesting. Although the spectrum is invariant, e.g. comparing $A=0$
and $A=2\pi/L$, the eigenmodes shift into each other,
$\j_{L,R}(p)\ra\j_{L,R}(p-2\pi/L)$, except for the last modes
with minimum momentum $p_{\rm min} = -\pi/2 + \pi/L$ which turn
into the $p_{\rm max} = \pi/2-\pi/L$ modes while flipping
chirality and flavor. This follows from the transformation
properties (\ref{lgt}) of the eigenmodes. The Dirac sea vacua
have all positive momentum $L$-modes and negative momentum
$R$-modes filled, such that $|0\rag_l\equiv|0,A=2\pi l/L\rag$
transforms under large gauge transformations
$\Om_k(x)=\exp(i2\pi x k/L)$ as
\be
\Om_k |0\rag_l = |0\rag_{k+l}, \label{lgtvac}
\ee
assuming a suitable choice of phases. The action of
$\widetilde{Q_5}$ on these states is given by
\be
\widetilde{Q_5} |0\rag_k = 2k |0\rag_k, \label{Qtildevac}
\ee
which also expresses the non-gauge invariance of this operator.

The vector QED$_2$ ground state energy $E_0(A)$ is periodic in $A$
with period $2\pi/L$, which expresses invariance under large gauge
transformations. In the continuum limit it reduces to the quadratic
form (\ref{E0}). The subtle regularization effects in the spectrum
away from the scaling region concord with gauge invariance and the $A$
dependence of $E_0$ is a surface effect of the Dirac sea
(cf. eq. (\ref{SURFACE})). In physical units, the relative
discretization errors in the $A$ dependence are of order $a^2 A^2$.
In the infinite volume limit the ground state energy density ${\cal
E}_0$ defined by
\be
{\cal E}_0(A) = \lim_{L\ra\infty} \frac{E_0(A)}{L},
\ee
becomes independent of $A$, ${\cal E}_0(A)={\cal E}_0(0)$, since the
variation in $E_0(A)$ is of order $1/L$ (cf. (\ref{E0})).

\begin{figure}[htb]
\epsfxsize=17cm
\centerline{ \epsffile[30 275 585 570]{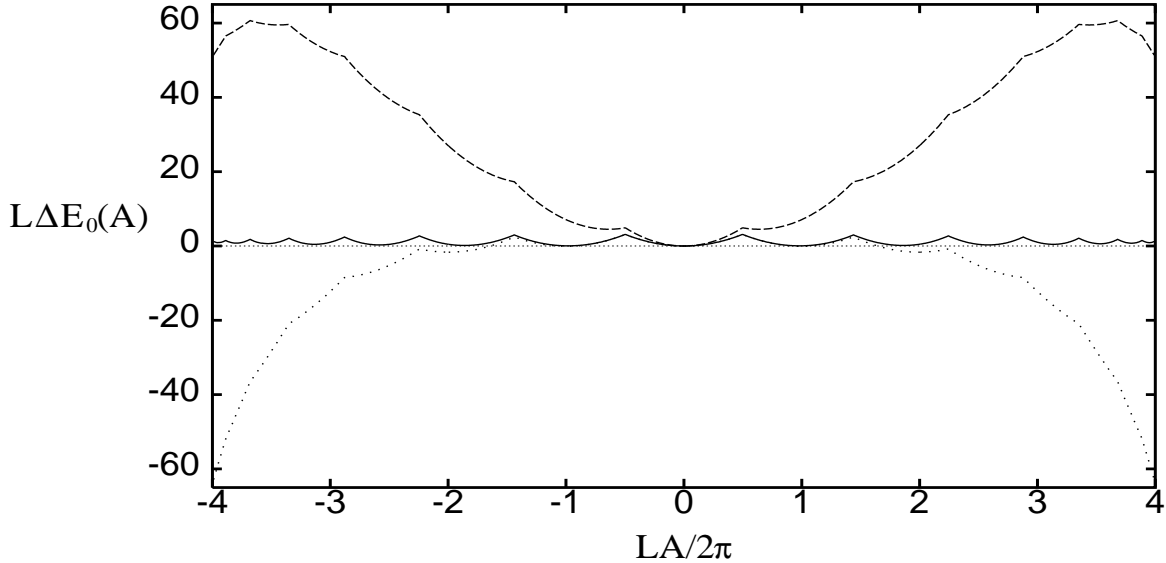}}
\caption{{\em The scaled energy difference
$L\Delta E_0 = L (E_0(A)-E_0(0))$ for the axial model as a function of
$ AL/2\pi$, on a lattice with $N=32$. Shown are the unrenormalized
energy: {\em dashed}; with perturbative counterterm $-\t L^2 A^2/2$
added: {\em dotted}; with $L^2({\cal E}_0(A) - {\cal E}_0(0))$
subtracted: {\em solid}.}}
\label{VACUUM}
\end{figure}

{}For the axial model fig.~\ref{FLOW} shows that the breaking of gauge
invariance has the effect that the spectra describing the vacua at
$A=2\pi/L$ and $A=0$ differ from each other, especially for high
momentum modes which are far away from the surface of the Dirac sea.
The regularization dependence of the axial spectrum is also a subtle
effect, which is just right so that the quadratic counterterm can
restore gauge invariance to the ground state energy in the scaling
region. In fig.~\ref{VACUUM} we have plotted the dimensionless
function $L(E_0(A)-E_0(0))$, without the counterterm, with the
counterterm $-\tau L^2 A^2/2$, and with a more refined subtraction in
terms of the energy density of the ground state, $-L^2{\cal
E}_0(A)$. The latter subtraction is equivalent to the perturbative
counterterm in the scaling region since ${\cal E}_0(A)-{\cal E}_0(0)=
\t A^2/2 + O(A^4)$ (cf. appendix A), but leads clearly to better
overall behavior for the renormalized ground state energy. (The size
of the scaling region will be discussed shortly.)

\begin{figure}[htb]
\epsfxsize=17cm
   \centerline{\epsffile[30 275 585 570]{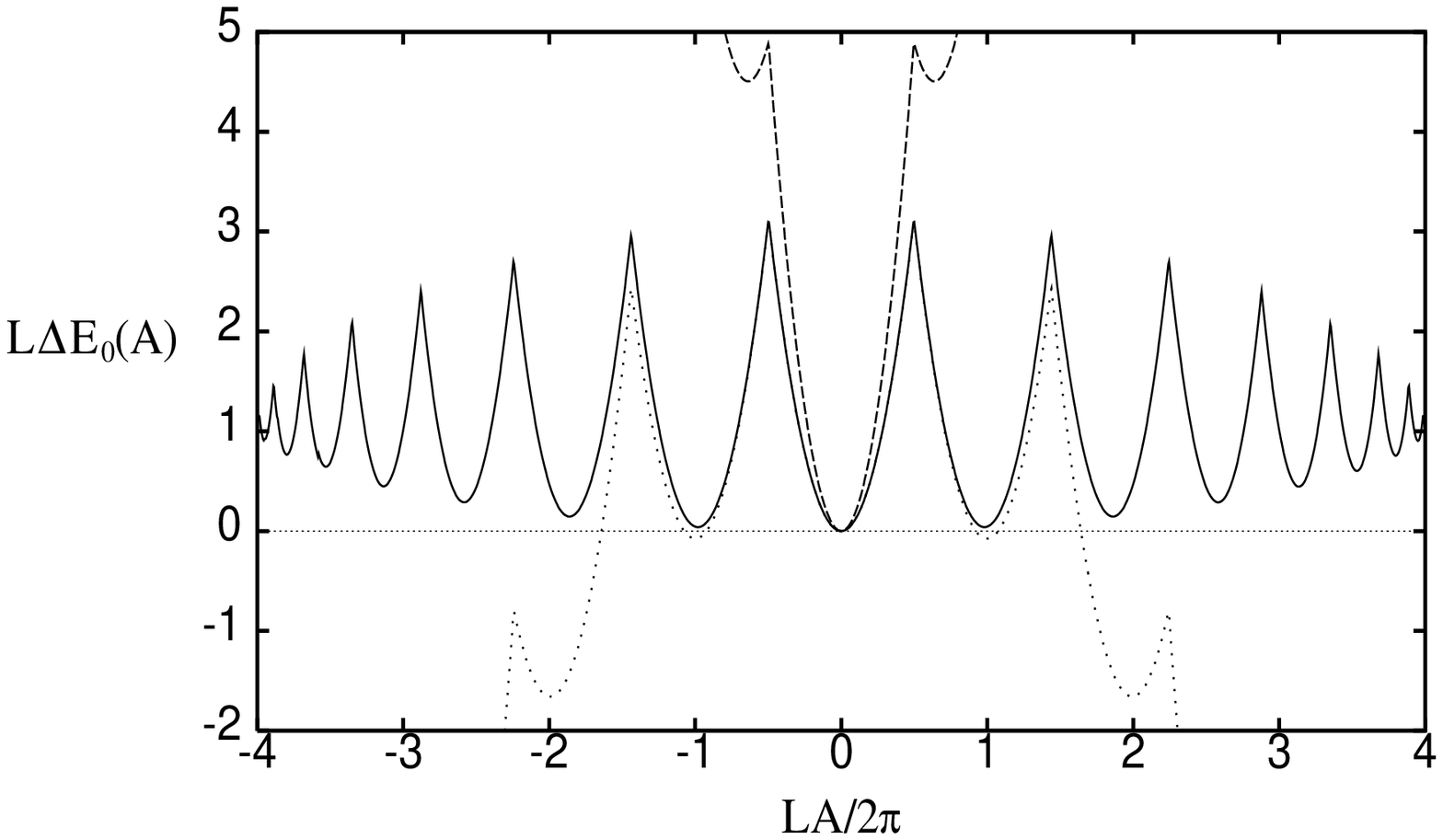}}
\caption{{\em As in fig. \protect \ref{VACUUM}, on a smaller
scale ({\em dashed}: unrenormalized energy; {\em dotted}: with
perturbative counterterm $-\t L^2 A^2/2$ added; {\em solid}: with
$L^2({\cal E}_0(A) - {\cal E}_0(0))$ subtracted).}}
\label{VACUUM2}
\end{figure}

{}Fig. \ref{VACUUM2} shows a close-up. We clearly see how the
counterterm restores gauge invariance in the scaling region, which for
a lattice of 32 sites appears to be given by $|LA|/2\pi\aleq 1$. The
cusps at $LA/2\pi = \pm 1/2$ correspond to the energy barrier between
the vacua $E_0(\pi/L)-E_0(0)=\pi/L$ (for two flavors). In the region
$-1
\leq LA/2\pi \leq 1$ containing the first three minima, the
vector and axial ground state energies are very close.

Notice that the abscissas in figs.   \ref{VACUUM} and
\ref{VACUUM2} are
limited to $|A|<\pi/4$; the reason is that at $A=\pm\pi/4$ (up to
corrections of order $1/L$) even the lowest/highest lying mode has
been pulled out/pushed into the Dirac sea: $\e_{\c}(p,A) = 0$ for
$p=\mp\pi/2$, $A=\pm \pi/4$. When the Dirac sea evaporates $(A\approx
\pi/4)$ or the Dirac sky condenses $(A\approx -\pi/4)$,
the physics clearly ceases to make sense, and we have to restrict the
allowed values of $A$ in the axial model to $|A|<\pi/4$.

We also see from figs. \ref{VACUUM} and \ref{VACUUM2} that the
quadratic counterterm is much too strong in the large field region,
outside the scaling region. With dynamical gauge fields it would be
bad to have the minima of the classical energy lying outside the
scaling region and away from $A=0$. To bring the minimum of the energy
into the scaling region we must add suitable higher order terms to the
action, e.g. $-\int d^2x\, \t' a^2 (A_{\mu}A_{\mu})^2$, with $\t'
>0$. In the static case the best choice is perhaps to subtract $L{\cal
E}_0(A)$ from the energy.

We have checked numerically using a variety of lattice sizes that for
given $LA$, the relative discretization errors are again $O(a^2
A^2)$. However, in contrast to the vector case, convergence is not
periodic in $LA$.  We can estimate the size of the scaling region to a
given accuracy on a spatial lattice of $N$ sites. To this end we
restore the lattice distance $a$, $L=Na$, and write $E_0(A)/L = a^{-2}
f_N(aA)$, ${\cal E}_0(A) = a^{-2} f_{\infty} (aA)$, where $E_0$ does
not include the counterterm. Figs. \ref{VACUUM} and \ref{VACUUM2} show
that $L\Delta E_0
\equiv L(E_0(A)-E_0(0)) = N^2 (f_N(aA) - f_N(0))$ behaves like
$L^2({\cal E}_0(A) - {\cal E}_0(0))= N^2
(f_{\infty}(aA)-f_{\infty}(0))$, modulated with `oscillations' which
in the scaling region turn into the physical $L\Delta E_0$ (i.e.
the quadratic $(LA)^2/\pi\;\,{\rm mod}\;2\pi$ potential).
Since $f_{\infty}(aA)-f_{\infty}(0)=
\half \t (aA)^2 + \t' (aA)^4 + \cdots$, we expect that
upon subtracting the counterterm,
$L\Delta E_0-\half\t (LA)^2$ equals the physical $L\Delta E_0$ up to terms of
order $N^2 (aA)^4 = N^{-2}(LA)^4$. It follows that to given accuracy,
$LA$ can be at most of order
$\sqrt{N}$. Hence the scaling region for $LA$ grows as $\sqrt{N}$.

In fig. 5 we display this scaling region for a lattice of
size $N=256$. The nonperturbatively renormalized ground state energy
$L\Delta E_0 - L^2\Delta {\cal E}_0$ of the axial model (cf. fig. 4)
is plotted superimposed on the gauge invariant (periodic) potential of
the vector model. One can clearly see that several axial vacua near
$A=0$ are degenerate and indistinguishable from the gauge invariant
vector case, with violations of gauge invariance setting in at
$|C| \ageq 5$.

In the continuum limit $N\ra\infty$ we recover the
equivalence with the vector model. The scaling region then
contains an arbitrarily large number of vacua $|0\rag_k$ which
transform into each other under large gauge transformations as in
(\ref{lgtvac}), with the conserved fermion charge acting as
\be
Q|0\rag_k = -2k|0\rag_k,
\ee
similar to (\ref{Qtildevac}). Clearly, the conserved $Q$ is not
gauge invariant; it is the analogue of $-\widetilde{Q_5}$ in
the vector model.

\begin{figure}[htb]
\epsfxsize=16cm
   \centerline{\epsffile[30 275 575 600]{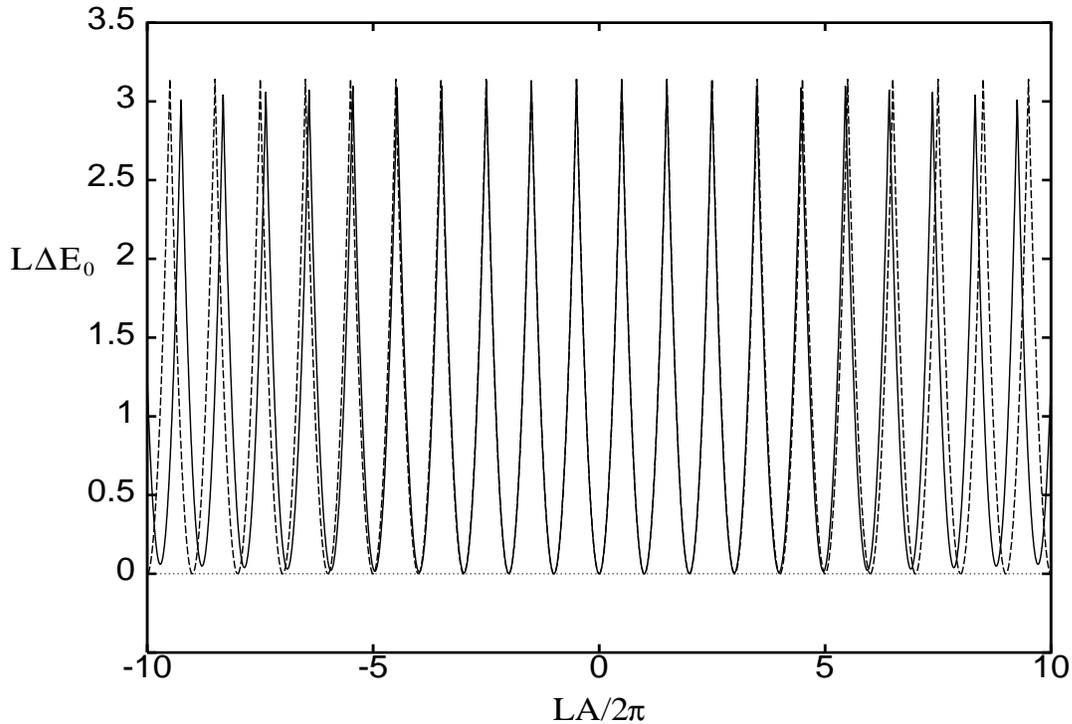}}
\caption{{\em Scaling region for a lattice of $N=256$ sites:
the renormalized vacuum energy of axial QED is represented by the
solid line, that of vector QED is dashed.}}
\label{VACUUM5}
\end{figure}

\section{Discussion}

We have shown using simple models in two dimensions,
from spectral flow in the external field approximation that fermion
creation and annihilation is possible even when fermion number is
exactly conserved. The transition rate is determined by the usual
anomaly formula, as if the fermion current had an anomalous
divergence. We see no reason why the same resolution of the electroweak
U(1) problem would not apply in four dimensions.

Furthermore, our study of the ground state energy of the axial
model reveals that a standard perturbative mass counterterm for the
gauge field, which was needed for restoration of gauge invariance, is
insufficient nonperturbatively. Additional terms are needed
(`oversubtraction') to get the global minimum of the ground state
energy in the desired scaling region. In the vector model, the
elegant way in which the staggered fermion method is able to give a
bottom to the Dirac sea \cite{AmGr83} in a gauge invariant way (thereby
circumventing the `infinite hotel' explanation of the anomaly
\cite{NN}) is also worth noting.

To complete our arguments, they have to be extended to the full theory
with dynamical gauge fields. We also have to recover zero modes,
fermion number violation in correlation functions, etc..
Since we
used continuous real time in the spectral flow arguments this may
not be a problem. However, with discrete time e.g. in the
euclidean formulation exact zero modes are non-generic and the
consequences of this will have to be understood (for analogous
questions in QCD see ref. \cite{SmVi88,Vi88}). These questions lie
outside the scope of our
present inquiry. Nonetheless we shall briefly outline some expectations for
the full theory with dynamical gauge fields.

Consider first the vector theory as defined by the lattice regulated
euclidean path integral in the compact formulation, in which the
dynamical gauge field variables are phase factors $U_{\mu\,
x\t}=\exp(-iA_{\mu\, x\t})$ with $A_{\mu\, x\t}\in(-\pi,\pi]$, in
lattice units. The path integral is then well defined without gauge
fixing and its Hilbert space interpretation leads naturally to
temporal gauge quantization, in which the integration over the time
components $U_{2\, x\t}$ of the gauge field provides the projection
onto the gauge invariant subspace, the implementation of `Gauss's
law'.  In the previous sections we used the Coulomb gauge however, so
let us fix to this gauge: $U_{1\,x\t} = U_{\t} = \exp(-iA_{\t})$
independent of $x$. In this case the integration over the time
components $U_{2\,x\t}$ in the path integral can be split into a
summation over discrete large gauge transformations $\Om_{k_{\t}} =
\exp(i2\pi k_{\t}x/L)$ plus fluctuations. The summation $P_0 = \sum_k
\Om_k$ over the large gauge transformations leads again to a
projection onto the gauge invariant subspace whereas the integration
over fluctuations leads to the Coulomb interaction. This
interpretation of the integration over $U_{2\,x\t}$ is appropriate in
the scaling region.

To compare with analysis in the continuum and also with the axial
theory to be discussed below, let us assume that in the scaling
region there is a $\widetilde{Q_5}$ which commutes with the
hamiltonian (but not with $P_0$). This is not entirely obvious,  since
with fluctuating gauge fields the spin-flavor interpretation of
staggered fermions emerges only in the scaling region and even
though $A$ is constant, the local charge densities in the
Coulomb interaction can only be expected to commute with such a
$\widetilde{Q_5}$ in the scaling region.
In the scaling region the Hilbert space before projection
with $P_0$ may then be split into sectors characterized by the
eigenvalue of $\widetilde{Q_5}$, $\widetilde{Q_5} |\Psi \rag_k =
2k |\Psi\rag_k$, similar to the external gauge field case. Under
large gauge transformations these sectors transform into one
another, $\Om_k |\Psi\rag_l = |\Psi\rag_{k+l}$, and consequently
physical states $P_0 |\Psi \rag_k = \sum_l |\Psi\rag_l$ have no
well defined value of $\widetilde{Q_5}$.
In particular $\widetilde{Q_5}$ does not annihilate the ground
state $|0\rag$, which is like a $\th$-vacuum superposition with
$\th = 0$, $|0\rag = \sum_k |0\rag_k$.

The axial case is more difficult since we
have not given a complete nonperturbative construction of the model
with quantized gauge fields. The difficulty is the lack of gauge
invariance at the fermion regulator scale. However, as mentioned in
the Introduction, we have in mind an implementation which is gauge
invariant on the scale of the lattice distance for the gauge
field, which is much larger than the lattice distance for the
fermions. Thus we may expect to regain invariance under gauge
transformations which are smooth on the fermion regulator scale,
in particular large gauge transformations, as we have seen in the
scaling region of the external field model.

This leads us to the conclusion that also in the axial case with
dynamical gauge fields $|0\rag = \sum_k |0\rag_k$ as well. Here
however it is
fermion number $Q$, which is analogous to
$\widetilde{Q_5}$ in the vector case, that is not well defined
in the vacuum. Since the corresponding global U(1) symmetry is an
exact symmetry of both the action and the fermion measure in the
path integral, this suggests that the symmetry is broken
spontaneously via the $\theta$-vacuum: $Q|0\rag = -\sum_k
2k|0\rag_k \neq 0$.
This picture though, needs to be reconciled with the expectation
that spontaneous symmetry breaking occurs only in the infinite volume
limit and that spontaneous breaking of a continuous symmetry in two
dimensions is not possible. This will have to await the detailed
construction of the model with dynamical gauge fields.

Despite the simple nature of the models studied here, we hope the
above picture is a valid description of the physics,
also in similar four dimensional models where it mimics old
ideas on the spontaneous breaking of U(1)$_{\rm A}$ in continuum
treatments of QCD.
Our conserved vector current $j^{\mu}$ in the axial QED$_2$ model is
similar to the analogous axial current in QCD, $\tilde{j}^{\mu}_5 =
i\jb\g^{\mu}\g_5\j - 2 n_f C^{\mu}$ ($n_f$ is the number of
flavors), which is conserved but not gauge invariant. The U(1)
symmetry generated by the conserved charge $\widetilde{Q}_5 = \int
d^3 x\,\tilde{j}^{0}_5$ is in this scenario supposed to be
spontaneously broken due to instanton---$\theta$-vacuum effects.
(There appears to be no universal agreement on this point
\cite{tH,Cr}); the U(1)$_{\rm A}$ problem in QCD can also be
solved without invoking a gauge variant conserved
$\tilde{j}^{\mu}_5$ \cite{Wi79}; for lattice implementations see
\cite{SmVi88}.)

To someone working exclusively in the continuum formulation,
where models are constructed from formal unregulated expressions
proceeding to well defined regulated expressions `along the way',
the gist of above remarks may appear so familiar that the question may
arise, what's new? We repeat however, that with a nonperturbative
regulator such as the lattice, models are well defined from the start
and we have to determine the fate of the exact global U(1) symmetry
and fermion number conservation. We have presented a resolution, at
least in the external field approximation.

\bigskip
\noindent {\bf Acknowledgements:} We would like to thank Jeroen C. Vink for
interesting discussions.  This research was supported by the
Stichting voor Fun\-da\-men\-teel On\-der\-zoek der Materie (FOM)
and by the DOE under contract DE-FG03-90ER40546.\\

\section*{Appendices}
\appendix
\section{Staggered fermion interpretation and ground state energy}
Various routes to the staggered fermion actions (\eq{CHIQED}) and
(\eq{CHIAX}) from the continuum actions (\eq{VCQED},\eq{AXQED}) have
been described in refs.~\cc{Sm88,BoSm94a,Sm92a,GoSm84}. Here we shall
briefly review the continuum interpretation of the staggered fermion
actions.

The interpretation is done in momentum space on a lattice of
$N_1 N_2$ points; for simplicity we use lattice units
$a=1$. The momenta in the Brillouin zone are parameterized as
$p_{\mu} + \bpi_{A\m}$, with $-\pi/2 < p_{\mu}< +\pi/2$ and $\bpi_A = \{
\bpi_0=(0,0), \bpi_1=(\pi,0),
\bpi_{12}=(\pi,\pi), \bpi_2=(0,\pi) \}$,
$A=1,\ldots,4$.  The regions of the Brillouin zone labeled by $A$
provide the spin-flavor components of the fermion field according to
$\ps^A_{p} = \sum_x\exp[-i(p + \bpi_A)x]\c_x$.  For the simple case of
constant $A_{\mu}$, the actions (\eq{CHIQED},\eq{CHIAX}) can then be
rewritten in the form
\cite{DoSm83,GoSm84}
\be
S_f = -\frac{4}{N_1 N_2} \widetilde{\sum_{p,\mu}}\, \jb_p\, i\G_{\mu}
\sin (p_{\mu}-A_{\mu}) \j_p, \label{MOMVEC}
\ee
for the vector case, and
\be
S_f = - \frac{4}{N_1 N_2}\widetilde{\sum_{p,\mu}}\, \jb_p\, \Big(i\G_{\mu}\sin
p_{\mu}\cos A_{\mu} + \sum_\nu\e_{\mu\nu}\G_{\nu} \cos p_{\nu} \sin
A_{\mu} \Big) \j_p,
\label{MOMAX}
\ee
for the axial case. Here the tilde on $\widetilde{\sum}_p$ indicates the
restricted momentum range given by $p_{\mu} = (n_{\mu} -1/2)
(2\pi/N_{\mu})$, $n_{\mu}=-N_{\mu}/4+1,\cdots,N_{\mu}/4$.
The $4\times 4$ real symmetric matrices $(\G_{\mu})_{AB}$ come out
\cite{DoSm83,GoSm84} naturally as tensor products of Pauli matrices
$\s_k$, and $\t_k$.  They commute with $4\times 4$ real symmetric
flavor matrices $\X_{\mu}$ which correspond to shift symmetries of the
staggered fermion action. The $\G_{\mu}$ and $\X_{\mu}$ are given
by
\be
\G_1 = \s_3\,,\;\;
\G_2 = \s_1\t_3\,,\;\;
\X_1 = \s_3\t_1\,,\;\;
\X_2= \t_3\,.
\ee
By a suitable unitary transformation the $\G$'s and $\X$'s can be
brought into block diagonal form, $\G_{\mu},\X_{\nu}\ra
\g_{\mu},\x_{\nu}$, with $\g_{1,2,5} = \s_{1,2,3}$, $\x_{1,2,5} =
\t_{1,2,3}$. The classical continuum limit corresponds to
small $p$ and $A$ where the sines and cosines take their limiting
forms.
In the general case of
non-constant gauge fields the classical continuum limit involves small
momenta and values of $A_{\mu x}$ as well.

The coefficient $\t$ of the counterterm of the axial model can be
obtained from the lattice vacuum polarization diagrams along the lines
in ref. \cite{BoSm94a}, which gives
\bea
\tau&=&-\left( \frac{1}{\pi}+I_1+I_2 \right)
\approx 0.3634\,,
\;\;\; \nnn
 I_2&=& -
\int_{-\pi/2}^{\pi/2} \frac{d^2 q}{\pi}\,
 \frac{ \sin^2 q_1 } { \sin^2 q_1 + \sin^2 q_2 } =-\frac{1}{2}
\,, \nonumber \\
I_1&=& \frac{1}{2}
\int_{-\pi/2}^{\pi/2} \frac{d^2 q}{\pi}\,
\frac{(\sin^2 q_2 -\sin^2 q_1) \cos^2 q_2
  }{\left(\sin^2 q_1 + \sin^2 q_2 \right)^2 }
 \approx -0.1817 \,.
\lb{INT}
\eea
This is
given in ref.~\cc{BoSm94a} for a somewhat different but equivalent
action. In this sense the action studied in
\cc{BoSm94a} has less gauge symmetry breaking than the `canonical
action' (\ref{CHIAX}) studied here (notice that the value in
ref.~\cc{BoSm94a} had been normalized to one flavor). Below we shall
obtain $\t$ again from the energy density of the ground state.

The above representations (\ref{MOMVEC}) and (\ref{MOMAX}) of the
actions are useful for calculating the ground state energy
$E_0=\lim_{N_2\ra\infty}~ -(1/N_2)\ln {\rm det}\,D$, where
$D_{xy}$ is the fermion matrix defined by $S_f = - \cb_x D_{xy}
\c_y$; we shall do so for the case $A_2=0$, $A_1=A$ considered in
this paper. In momentum space ${\rm det}\,D =
\tilde{\prod}_p {\rm det}\, D_{p}$, where $D_p = i\G_2\sin p_2 +
\cdots$ can be read off from (\ref{MOMVEC}) and (\ref{MOMAX}).
We write $D_p =\G_2 (i\sin p_2 + {\cal H}_p)$, where ${\cal H}_p$ is a
hermitian matrix. Since ${\rm det}\, \G_2=1$, the determinant of $D$
is given by the product of the eigenvalues of $(i\sin p_2 + {\cal
H}_p)$, which come in complex conjugate pairs. Hence, for the
evaluation of $\ln {\rm det}\, D$ we may use $\ln (z z^*) = \ln z +
\ln z^*$, with the principal value for the logarithm ($\ln z =
\ln |z| + i\, {\rm arg}\, z$, $-\pi < {\rm arg} z < \pi$).
{}For the vector theory we find
\bea
{\cal H}_p &=& \Gamma_5 \sin(p - A)\\
E_0 &=& - \widetilde{\sum_p} \int_{-\pi/2}^{\pi/2} \frac{dp_2}{\pi}
\sum_{\c} \ln (i\sin p_2 + \c \sin (p-A)),
\eea
and for the axial theory
\bea
{\cal H}_p &=& \Gamma_5 \cos A \sin p + \sin A \cos p_2\\
E_0 &=& - \widetilde{\sum_p} \int_{-\pi/2}^{\pi/2} \frac{dp_2}{\pi}
\sum_{\c}\ln (i\sin p_2 + \c \cos A \sin p + \sin A \cos p_2),
\eea
where we have let $N_2\ra\infty$, $p=p_1$, $N=N_1$, and the chirality
$\c=\pm 1$ is the eigenvalue of $\G_5$.

As an example we describe the calculation for the axial case. The
integrand has period $\pi$, so we may replace $\int_{-\pi/2}^{\pi/2}
dp_2/\pi \ra \int_{-\pi}^{\pi} dp_2/2\pi$.  The `fermion doubling'
introduced by this replacement is compensated by the factor 1/2.
Next we change variables to $z=\exp ip_2$ and use
\be
{\rm Re} \int_{|z|=1} \frac{dz}{2\pi i z} \ln(z-x) = \ln(|x|)\,
\th(|x|-1),
\label{formula}
\ee
where $x$ is real and $\th$ is a Heaviside step function,
$\th(x)=1$, $x>0$, $\th(x)=0$, $x<0$.
We factorize $i\sin p_2 +\c \cos A \sin p + \sin A \cos p_2 = (1+\sin
A)(z-z_+)(z-z_-)/2z$, with
\bea
z_{\pm} &=& \pm\exp(-\e_{\pm\c})\,,\;\;\;
\e_{\pm\c}=\pm\e_{0\c} + \ln\sqrt{(1+\sin A)/(1-\sin A)}\,,\label{EPAX}\\
\e_{0\c}&=&\ln (\c s+\sqrt{s^2 +1})\,,\;\;\;
\sinh \e_{0\c} = \c s\,,\;\;\;
s= {\rm sgn}(\cos A)\sin p. \label{EP0AX}
\eea
{}From the `inverse Wick rotation' $p_2=ip^0$ we see that $z_+$ and
$\e_+$ correspond to `normal fermion' excitations, whereas $p_2=ip^0 +
\pi$ shows that $z_-$ and $\e_-$ correspond to `doubler fermion'
excitations introduced by doubling the momentum interval. Furthermore,
the doublers have effectively opposite chirality,
$\e_{-,\c}=\e_{+,-\c}$. The step function in (\ref{formula})
allows only the negative $\e_{\pm\c}$ to contribute, as expected for the
Dirac sea. The axial ground state energy can be evaluated as
\be
E_0 = N\ln 2 - N\ln (1+\sin A) + 2 \widetilde{\sum_p}\sum_{\c}\,
\e_{\c}\,\th(-\e_{\c})\,,
\ee
where $\e_{\c}=\e_{+\c}$ depends on $A$ and $p$ according to
(\ref{EP0AX}).

The corresponding formulas for the vector case can be obtained from
the above by letting $A\ra 0$ and then $p\ra p-A$:
\bea
\e_{\c} &=& \ln (\c\sin(p-A) + \sqrt{\sin^2(p-A) + 1})\;\;\;
\sinh \e_{\c} = \c \sin(p-A)\;,\\
E_0 &=&N\ln2 + 2 \widetilde{\sum_p}\sum_{\c}\,
\e_{\c}\,\th(-\e_{\c})\,.
\eea
Using $\e_{-}(p,A) =\e_{+}(p+\pi,A)$ this can be rewritten as a
summation involving $\e_+$ only, over the whole periodic region
$p\in(-\pi,\pi)$.  This shows that $E_0(A)$ is periodic in $A$ with
period $2\pi/N$.  In the infinite volume limit the momentum summation
may be replaced by an integration over $p\in (-\pi,\pi)$, and by
$p$-translation invariance the energy density becomes independent of
$A$, as expected from gauge invariance.

Reintroducing the lattice distance $a$ by $A\ra aA$, $E_0\ra aE_0$,
$L=Na$, the continuum limit of the finite volume $E_0$ can be found by
expansion in $a$, taking $A$ in the first period about $A=0$. From
$\e_-(p,A) = \e_+(-p,-A)$ and the symmetry of the momentum summation,
only even powers of $A$ occur.  Using $\partial
\e_+/\partial A = -\cos(p-A) /\cosh \e_{+}$ and replacing afterwards
$\widetilde{\sum}_p$ by an integral for $a\ra 0$, gives
\bea
E_0(A)-E_0(0) &=& LA^2 \int^{0}_{-\pi/2} \frac{dp}{\pi}\,
\left[\frac{-\cos^2 p\, \sin p}{(1+\sin^2 p)^{3/2}}
+ \frac{-\sin p}{(1+\sin^2 p)^{1/2}}\right] + \cdots~~~~~~~~~
\label{TTT}\\
&=& LA^2 \int^{0}_{-\pi/2} \frac{dp}{\pi}\,
\frac{\partial}{\partial p}\frac{\cos p}{(1+\sin^2 p)^{1/2}}
+ \cdots
\label{SURFACE}\\
&=& \frac{1}{\pi} LA^2\, [1 + O(a^2 A^2)]\,.
\eea
The two terms in (\ref{TTT}) correspond to the `ordinary' and `leaf
diagram' terms in the vacuum polarization diagram, and the
differential identity in (\ref{SURFACE}) is essentially a Ward
identity \cite{KaSm81}. The factor two in comparison to the continuum
formula (\eq{E0}) is due to the two staggered flavors.

Returning to the axial case, for which $E_0$ is not gauge invariant
without the counterterm, we write $E_0 = N f_N(A)$ (in lattice units).
In the infinite volume limit the energy density $E_0/N$ approaches a
smooth function of $A$,
\bea
f_{\infty}(A) &=& \ln 2 - \ln(1+\sin A) + 4\int_{-\pi/2}^{\pi/2}
\frac{dp}{\pi}\,
\e_+(p,A)\th(-\e_+(p,A)) \nnn \\
&=&\ln 2 - \ln(1+\sin A) + \frac{4}{\pi}\int_{-\pi/4}^A d\th\,
\frac{\sqrt{\cos 2\th}}{\cos \th}\,\ln\sqrt{\frac{1+\sin
A}{1-\sin A} \; \frac{1+\sin\th}{1-\sin\th}}\;,\nonumber
\lb{finf}
\eea
where in the second line we used the substitution $\sin p =
\tan\th$ and assumed $A>0$. {}From $f_{\infty} (A) = \half\t A^2
+ O(A^4)$ we get an analytic expression for the coefficient of the
counterterm,
$\t = 1-2/\pi  = 0.363380..$.
\section{Transfer operator and energy spectrum in vector QED$_2$}
In this appendix we construct the transfer operator for vector
QED$_2$ and rederive from it the energy spectrum. To the path
integral a quantum mechanical Hilbert space is associated according to
the coherent state formalism \cite{ShTh82,Lu77}.
A rudimentary set of formulas for this purpose is given by
\bea
|a\rag &=&\exp(\ahd_i a_i)\noq\,,\;\;
\lag a|a\rag =\exp(a^+_i a_i)\,,\;\;
1=\prod_i \int da^+_i da_i \exp (- a^+_i a_i) |a\rag\lag a|\,,
\nonumber\\
\ah_i|a\rag &=& a_i|a\rag\,,\;\;\lag a|\ahd_i = a^+_i\lag
a|\,,\;\;
\lag a|\exp(\ahd_i M_{ij} \ah_j) |a\rag =
\exp[a^+_i (e^M)_{ij} a_j]\,. \label{RULES}
\eea
In these appendices we distinguish operators in Hilbert space by a
caret $\hat{\mbox{}}$.  The $\ah_i$ and $\ahd_i$, are a generic set of
fermionic creation and annihilation operators with the usual
anticommutation rules and `empty state' $\noq$, the $a_i$ and $a^+_i$
are `Grassmann variables', and $M$ is some matrix.

In the temporal gauge $U_{2,x}=1$, we can rewrite the vector action
(\eq{CHIQED}) in the form, using $\cb_x = \c^+_x \h_{2x}$,
\be
S=-\sum_{\t=0}^{N_2-1} [\half(\c^+_{\t} \c_{\t+1}-\c^+_{\t+1} \c_{\t})
+ \c^+_{\t} \CH_t \c_{\t} ]\,,
\ee
where we have used $(x_1,x_2) = (x,\t)$ and suppressed the
summations over $x$. The identification $\c_{x, N_2} = -\c_{x,0}$,
$\cb_{x,N_2} = -\cb_{x,0}$ takes into account the antiperiodic
boundary conditions. The staggered fermion Dirac hamiltonian $\CH$ is
given by
\be
\CH_{\t\;xy} = \half \;
(\h_{2\;x,\t} \; \h_{1\; x,\t} \; U_{1\;x,\t} \; \d_{y,x+1} +
\h_{2\;y,\t}  \; \h_{1\;y,\t} \; U_{1\;y,\t}^* \; \d_{x,y+1})
\;,\lb{HQED}
\ee
and is manifestly hermitian.

The time derivative terms $-\c^+_{\t} \c_{\t+1}+\c^+_{\t+1}
\c_{\t}$ are associated with the normalization factor
$\exp(a^+ a)$ and measure factor $\exp(-a^+ a)$ in the coherent state
formalism. This leads to the two time-slice construction described in
refs.~\cc{ShTh82,DoSm83,Sm91}, where we interpret the $\c$ fields
at odd time slices as `type $a$' and at even time
slices as `type $a^+$', writing
\bea
\c_{\t}&=& \sqrt{2}a_k\,,\;\;\;\c^+_{\t} = \sqrt{2} b_k\,,
\;\;\; \t=2k+1,\nonumber\\
\c_{\t}&=&\sqrt{2}b^+_k\,,\;\;\;\c^+_{\t} = \sqrt{2} a^+_k\,,
\;\;\; \t=2k, \label{CTOA}
\eea
where $k=0,\ldots, N_2/2-1$ denotes pairs of time slices.  Using these
new variables the partition function can be rewritten in the form
\bea
Z &=& \int \prod_{k=0}^{N_2/2-1} d\ad_k \, d a_k \, d\bd_k \, d b_k \;
\lag a_{k+1},b_{k+1}| \Th | a_{k}, b_{k} \rag\\
&=& \Tr \Th^{N_2/2}\,,
\eea
with
\be
\lag a_{k+1},b_{k+1}| \Th | a_{k}, b_{k} \rag     =
e^{ -2 N \ln 2 } \;\; e^{ \ad_{k+1} a_k + \bd_{k+1} b_k } \;\; e^{ -2
\ad_{k+1} \CH_{2k+2} \bd_{k+1} } \;\; e^{ -2 b_k \CH_{2k+1} a_k } \;.
\lb{TT}
\ee
 Recall that $N=N_1$ is the total number of spatial sites.
For the transfer operator $\Th$ associated with the matrix element
$\lag a_{k+1},b_{k+1}|\Th| a_{k}, b_{k} \rag$ we obtain the expression
\be
\Th = e^{  -2 N \ln 2  } \;\;
e^{ -2 \ahd \CH \bhd } \;\; e^{ -2 \bh \CH \ah } \;.  \lb{TO}
\ee
The staggered transfer operator can be written as the square of the
shift operator in the time direction
\cite{GoSm85,Altea93},
but we do not need this here.

We shall now determine the eigenvalues of $\Th$ for the case
$U_{1\;x,\t} = \exp(-iA)$ needed for the spectral flow ($A$ may be
considered constant in this calculation, as it is independent of the
euclidean time $\t$). First we consider the Dirac hamiltonian
$\CH$. Similar to the two dimensional case in (\ref{MOMVEC}),
the
spatial Fourier transform brings $\CH$ in a more intelligible form
\bea
\sum_{x,y} e^{-i(p+\pi_A)x+i(q+\pi_B)y}\; \CH_{xy} &=&
(\s_2)_{AB}\;\sin(p-A)\; N \;\d_{p,q}\;, \\
\sum_x e^{-i(p+\pi_A)x} \, \ah_x &=& \ah_{Ap}\;,\;\;\;{\rm etc.}\\
\bh\CH\ah &=& \frac{2}{N}\widetilde{\sum_p}
\half\bh_p\s_2\sin(p-A)\ah_p\;,\;\;\;
{\rm etc.}
\eea
Here the indices $A,B$ take only two values, $\pi_A=\pi_0,
\pi_1$, and in $\s_2$ we recognize the upper block of
$\G_5=\s_2\t_3$. We will recover the complete $\G_5$ below.  The
eigenvalues of $\CH$ are given by $\l=\pm\sin(p-A)$, where $\pm 1$ is
the eigenvalue of $\s_2$.
The transfer operator reduces to $\prod_{\rm modes}
\exp(-2\ln2) \exp(-2\ahd\l\bhd)\exp(-2\bh\l\ah)$.

For the moment we continue the calculation for a single mode.  We seek
linear combinations of $\ah,\bhd$ and $\ahd,\bh$ which have simple
commutation relations with $\Th$. This leads to a $2\times 2$ real
symmetric eigensystem for the coefficients of the linear
combinations. From
\be
\left(\begin{array}{c}\ah\\ \bhd\end{array}\right) =
V' \left(\begin{array}{c} \jh_+\\ \jh_-\end{array}\right)\,,\;\;\;
\ee
it is straightforward to check that
\bea
\Th \jh &=&
\left(\begin{array}{cc} e^{2\o'}&0\\0&e^{-2\o'}\end{array}\right)\jh\Th
\equiv e^{2\o'\t_3}\jh\,,\\
\o' &=& \ln(\l + \sqrt{\l^2+1})\,,\;\;\; \sinh \o' = \l\,,
\eea
with the orthogonal matrix $V'$ given by
\be
V' = \frac{1}{\sqrt{2\cosh
\o'}}\left(\begin{array}{cc}e^{-\o'/2}&e^{\o'/2}\\
-e^{\o'/2}&e^{-\o'/2}\end{array}\right)\,.
\ee
We can now retrace our steps and restore the matrix $\s_2$ to
the eigenvalues of ${\cal H}$ in this basis:
$\l\ra\s_2\sin(p-A)$, $\o'\ra\s_2\o$, turning $\jh_{\pm}$ into two
component fields on which the $\s$'s  act, hence obtaining four
component fields $\jh$ on which both the $\s$'s and $\t$'s
act. Then the $2\times 2$ orthogonal $V'$ gets replaced by a
$4\times 4$ unitary matrix $V$,
\bea
\left(\begin{array}{c}\ah\\ \bhd\end{array}\right) &=&
V \jh,\\
\Th\jh &=&e^{2\o\s_2\t_3} \jh \Th\,,\;\;\;
\sinh \o = \sin(p-A)\,,\label{omegapA}\\
V&=&\frac{1}{\sqrt{2\cosh \o}}(e^{-\o\s_2/2} + i\t_2
e^{\o\s_2/2})\,, \label{TjjT}
\eea
and we have recovered $\G_5 = \s_2\t_3$.

Defining the no-quantum state $\noq$ by $\ah\noq=\bhd\noq = 0$, it
follows that $\jh\noq = 0$ and the eigenstates of the transfer
operator are obtained by repeated application of $\jhd$ on $\noq$,
\bea
\Th\jhd\noq &=& e^{-2\o\G_5}\jhd \Th\noq\,,\\
\Th \noq &=& e^{-2N\ln2}\noq\,.
\eea
The mode eigenvalues are obviously given by $\c\o$, with
$\c$ the eigenvalue of $\G_5$.

Under a large gauge transformation
\be
\Ohd\ah_x\Oh = e^{i2\pi x/N} \ah_x\,,\;\;\;,
\Ohd\bhd_x\Oh = e^{i2\pi x/N} \bhd_x\,,
\ee
the $\jh_p(A)$ transform as
\bea
\Ohd\jh_p(A+2\pi/N)\Oh&=&\jh_{p-2\pi/N}(A),\;\;\;\;\;
p=p_{\rm min}+ 2\pi/N,\cdots,p_{\rm max},\nonumber\\
\Ohd\jh_{p_{\rm min}}(A+2\pi/N)\Oh&=& \G_2\Xi_2 \jh_{p_{\rm max}}(A),
\label{lgt}
\eea
where we used $\sg_1 =\G_2\Xi_2$ and $p_{\rm min} = -p_{\rm max}
= -\pi/2 +\pi/N$. We see that the minimum momentum mode flips its
chirality $\c$ and flavor $\xi$ (eigenvalue of $\Xi_5$) as it changes into
the maximum momentum mode,
\be
\Ohd\jh_{p_{\rm min}}^{\c,\xi}(A+2\pi/N)\Oh=
(-1)^{(\c+\xi)/2}\, \jh_{p_{\rm max}}^{-\c,-\xi} (A) \,.
\ee
The eigenvalue of $\G_5\Xi_5$ remains the same.

We shall now obtain operator expressions for the charges $Q$, $Q_{\e}$
and $Q_5$. Since fermion number $Q=-i\sum_x j_{2\; x,\t}$ is time
independent we choose $\t=0 = {\rm even}$, and associate with $Q=\half
(\c^+_{\t}U_{2,\tau}\c_{\t+1} + \c^+_{\t+1}U_{2,\tau}^* \c_{\t})
= a^+ a + b b^+ $
the operator
\be
\Qh = \ahd\ah -\bhd\bh = \ahd\ah +\bh\bhd - N =
\frac{2}{N}\widetilde{\sum_p} \half\jhd_p\jh_p - N\,.\label{QOP}
\ee
This has the expected form (\ref{QFORM}) for two flavors. The other
conserved U(1) charge $Q_{\e}= -i\sum_x j^{\e}_{2\; x,t}= \sum_x
\half(-\e_x\c^+_{\t}\c_{\t+1} + \e_x\c^+_{\t+1}\c_{\t})$ is also
time independent (recall $\e_x = (-1)^{x_1+x_2}$).
Choosing $\t=0$ again leads to
\bea
\hat{Q}_{\e} &=& -\frac{2}{N}\widetilde{\sum_p}
\half(\ahd_p\s_1\ah_p + \bhd_p\s_1\bh_p) =
-\frac{2}{N}\widetilde{\sum_p} \half\jhd_p V^{\dagger}\s_1\t_3 V
\jh_p\nonumber\\
&=&\frac{2}{N}\widetilde{\sum_p} \half \jhd_p
\G_5\X_5\jh_p\,,
\eea
with $\G_5\X_5=-\s_1\t_1$ the flavor non-singlet axial generator
corresponding to $\e_x$.

Although the chiralities $\c$ appear naturally in the mode spectrum of
the staggered formulation, an axial current $j_{5\mu}$ or charge
operator $Q_5$ is somewhat extraneous to the formulation. A natural
candidate is given by $j_{5\mu} = -i\e_{\mu\nu} j_{\nu}$, $Q_{5} =
\sum_x j_{1\;x,\t}$. However, for a reasonable interpretation in
staggered fermion theory, operators usually have to involve some time
average as well as space average. This leads us to investigate
$Q_{5}\equiv \half \sum_x (j_{1\;x,\t} + j_{1\;x,\t+1})$ at $\t=0$.
We find
\bea
\Qh_5 &=& -\frac{2}{N}\widetilde{\sum_p}
\frac{1}{2}(\ahd_p \s_2\bhd_p + \bh\s_2\ah) \cos(p-A),\nonumber\\
&=& -\frac{2}{N}\widetilde{\sum_p}\half\jhd_p V^{\dagger} \t_1\s_2 V
\jh_p\cos(p-A)\,,\label{Q5}\\
&=& \frac{2}{N}\widetilde{\sum_p} \half\jhd_p \left[\frac{\G_5}{\cosh \om}
 + \G_1\Xi_1\tanh \om \right]\jh_p\cos(p-A), \nonumber
\eea
where $\G_1\Xi_1 = \ta_1$; note that $\om$ depends on $p-A$ according to
eq. (\ref{omegapA}).
In the scaling region $\cos(p-A)\ra 1$, $\om\ra 0$ and the
rather ugly looking $\Qh_5$ reduces to the continuum form.
Away from the scaling region there is a reduction of
the `strength of $\G_5$' in $Q_5$ and flavor symmetry breaking.

The operators $\Qh$, $\Qh_{\ep}$ and $\Qh_5$ are invariant under
the large gauge transformations (\ref{lgt}), as is obvious from
their original gauge invariant expression in terms of the
staggered fermion fields.
We can define similarly a {\em conserved} chiral charge operator
$\hat{\widetilde{Q_5}}$ which commutes with the transfer operator (\eq{TO})
but is not gauge invariant under the large gauge transformations
(\ref{lgt}),
\be
\hat{\widetilde{Q_5}} = \frac{2}{N}\widetilde{\sum_p} \half\jhd_p
\G_5\jh_p.
\label{Qtilde}
\ee
The chiralities $\c=\pm 1$ of the modes $\j_{R,L} = \half(1\pm
\G_5)\j$ are the eigenvalues of this operator
$\hat{\widetilde{Q_5}}$.

\section{Transfer operator and energy spectrum in axial QED$_2$}
To calculate the transfer operator and its eigenvalue spectrum in the
axial-vector model we follow the same steps as in the previous
section, and will therefore only point out the differences in the
calculation as compared to the vector case. Using the definitions
(\eq{CTOA}), we find for the action (\eq{CHIAX}) in terms of the
fields $\c_{\t}$ and $\c^+_{\t}$, the expression
\bea
S&=&-\sum_{\t=0}^{N_2-1} [\half(\a\,\c^+_{\t}
\c_{\t+1}-\b\,\c^+_{\t+1} \c_{\t}) + \c^+_{\t} \CHT_{\t} \c_{\t} ]\,,
\lb{AXX} \\
\a&=&1+\sin A\,,\;\;\;\b=1-\sin A\,.
\eea
The matrix $\CHT_{\t\;xy}$ in (\eq{AXX}) is given by the same
expression as in (\eq{HQED}), except that $U_{1\;x,\t}$ has to be
replaced by the real form $\cos A$. With the same assignment of the
variables $a_k,\ldots,b_k^+$ to the $\c_{\t}$ and $\c^+_{\t}$ fields,
as in (\ref{CTOA}), and after the rescaling $a_k~(\ad_k) \ra
a_k/\sqrt{\a}~ (\ad_k/\sqrt{\a})$ and $b_k~(\bd_k) \ra  b_k/\sqrt{\b}~
(\bd_k/\sqrt{\b})$ we find that the matrix element of the transfer
operator is given by
\bea
&& \!\!\!\!\!\!\!\!\!\!\!\!  \lag a_{k+1},b_{k+1}| \Th | a_{k}, b_{k}
\rag = e^{ -2 N \ln 2 } \;\; e^{ N \ln (\a \b) } \nnn &&
\!\!\!\!\!\!\!\!\!\!\!\!
\times e^{  -2 \ad_{k+1} (\CHT_{2k+2}/\sqrt{\a\b}) \bd_{k+1}  }  \;\;
e^{(\b/\a)\ad_{k+1} a_k + (\a/\b) \bd_{k+1} b_k } \;\; e^{ -2 b_k
(\CHT_{2k+1}/\sqrt{\a\b}) a_k } \;.  \lb{TTAX}
\eea
The second exponential factor in (\eq{TTAX}) arises from the jacobian
in the above rescaling of the Grassmann measure:
$d(a_k/\sqrt{\a})=\sqrt{\a}\, d \a_k$, etc.  Using the rules
(\ref{RULES}) for transcribing the Grassmann algebra matrix elements
to their associated Fock space operators we obtain from (\eq{TTAX})
\bea
&&\!\!\!\!\!\!\! \Th = e^{ -2 N \ln 2 } \;\; e^{ +N \ln (\a \b) } \;\;
e^{ -2 \ahd \CH_0 \bhd } \;\; e^{ \ln(\b/\a) \; (\ahd \ah - \bhd \bh)
} \;\; e^{ -2 \bh \CH_0 \ah } \;, \lb{TOAX} \\ &&\!\!\!\!\!\!\!
\CH_{0\;\t\;xy} = {\rm sgn}\,(\cos A) \half \;
(\h_{2\;x,\t} \; \h_{1\;x,\t} \; \d_{y,x+1} + \h_{2\;y,\t} \;
\h_{1\;y,\t} \; \d_{x,y+1})
\;. \lb{DEFAAX}
\eea
{}For ${\rm sgn}\, \cos A >0$, which is satisfied because we require
$|A|<\pi/4$ (cf. sect. 4), the matrix $\CH_0$ is just the free
staggered fermion Dirac operator. In the factor $\ahd \ah - \bhd
\bh$ we recognize the charge operator $\Qh$ of (\ref{QOP}). The factor
$\exp[\ln(\a/\b)\Qh]$ has simple commutation relations with the fields
$\ah,\bhd$ and $\ahd,\bh$ and the analogue of (\ref{TjjT}) in this
axial case is given by
\bea
\Th\jh &=&e^{\ln(\a/\b) + 2\o_0\s_2\t_3}\; \jh \Th\,,\;\;\;
\sinh \o_0 = \sin p\,,\label{TjjTAX}\\
V&=&\frac{1}{\sqrt{2\cosh \o_0}}(e^{-\o_0\s_2/2} + i\t_2
e^{\o_0\s_2/2})\,. \label{OAX}
\eea
The interpretation of the operators $\jh$, $\jhd$ follows that of the
vector case.  The energy spectrum following from (\ref{TjjTAX}) is the
same as in (\ref{EPAX}).

%
%

\end{document}